# Identifying Differences in Diagnostic Skills between Physics Students: Students' Self-Diagnostic Performance Given Alternative Scaffolding

E.Cohen[2], A. Mason[1], C. Singh[1] and E. Yerushalmi[2]

[1]*Department of Physics and Astronomy, University of Pittsburgh, Pittsburgh, PA 15213, USA*
[2]*Department of Science Teaching, Weizmann Institute of Science, Rehovot, Israel*

**Abstract.** "Self-diagnosis tasks" aim at fostering diagnostic behavior by explicitly requiring students to present diagnosis as part of the activity of reviewing their problem solutions. We have been investigating the extent to which introductory physics students can diagnose their own mistakes when explicitly asked to do so with different levels of scaffolding support provided to them. In our study in an introductory physics class with more than 200 students, the recitation classes were split into three different experimental groups in which different levels of guidance were provided for performing the self-diagnosis activities. We present our findings that students' performance was far from perfect. However, differences in the scaffolding in the three experimental groups (i.e. providing a correct solution and a self-diagnosis rubric) noticeably affected the resulting diagnosis.

## INTRODUCTION

In the companion paper [1], we laid out a rubric to evaluate self-diagnosis in the context of scaffolded diagnostic tasks in which students are encouraged to self-diagnose their problem solutions by providing credit and a time slot in class for self-diagnosis.

Requiring students to self-diagnose their solutions is motivated by several assertions:

1) Problem solving is a learning opportunity; solvers need to reflect to learn from their solution and mistakes. Yet, many students are not able to effectively take advantage of this opportunity, either because they lack reflective practice and skills or because they don't have a perception of problem solving as a reflective process [2, 3].

2) We expect students to be able to reflect on their solution if deliberately prompted to do so, and if they are provided with appropriate scaffolding aligned with cognitive apprenticeship [4], namely, incorporating three elements: (a) modeling (in our context this involved the TA demonstrating how he would perform the diagnostic task on a mistaken solution he provides the students), (b) coaching (in our context this involved providing students opportunity to perform a diagnostic task, reflecting and diagnosing their solution with guidance). Clearly the coaching was minimal, taking into account the time constraints of TAs in a large college classroom. (c) "fading" (i.e. decreasing the support and feedback.) (d) students' prior knowledge and approaches will influence the level of diagnosis they will perform. In terms of the experiment, one might expect that students who solved the problem better initially would also diagnose better, and, that minimal guidance will better serve the high-achievers, while the worked out examples will help the lower-achievers more.

While the constraints of introductory courses imply minimum intervention (minimum modeling and feedback for the self-diagnosis task), it might be that such intervention would not be sufficient. Namely, one session of modeling, and almost no feedback on the diagnosis, would not be enough to allow students who don't regularly engage in reflecting on their solution to do so in a meaningful way.

In this paper we present findings regarding students' performance on three self-diagnosis tasks. The instructions and resources students receive differ in the scaffolding provided from thorough to minimal scaffolding. In the first self-diagnosis task, the instructor outlines the correct solution, and students fill in a self-diagnosis rubric including problem solving steps commonly found in different problem solving strategies. In the second task, they compare their own solution with a worked out example, and in the third they receive merely the final result with minimal guidance. A more complete description of the diagnostic tasks and the experimental layout is featured in former work [5]. In a subsequent paper [6] we will present the effect of the diagnostic tasks on solving isomorphic and far transfer problems. In particular we will answer the following questions:

1) *What are students able to diagnose in the different self-diagnosis groups?*
2) *How does students' performance of self-diagnosis depend upon their prior knowledge?*

## EXPERIMENTAL SETUP & SAMPLE

The study involved an introductory algebra based course for pre-meds (N~200), one instructor and two teaching assistants. TA classrooms were distributed into control groups and three self-diagnosis treatments groups each carried out a different self-diagnosis task. In all treatment groups, students first solved a quiz problem, and in the next training session they were asked to circle mistakes in their photocopied solutions and explain what they did wrong:

**TABLE 1.** Distribution of groups - self-diagnosis tasks

| | Self-diagnosis tasks | | |
|---|---|---|---|
| | Outline + Rubric | Worked out example | minimal guidance |
| Resources | Instructor outline the correct solution, Students fill in a self-diagnosis rubric | Instructor provides written sample solution | Students can use their notes + text books |
| | 1 group (B), 31 students | 1 group (C), 28 students | 1 group (D), 25 students |

## FINDINGS

*The first question we address is: what are students able to diagnose in the different self-diagnosis group?*

To compare the groups we performed ANCOVA analysis taking the quiz physics grade as a covariate (after checking homogeneity of slopes).

We took the $1^{st}$ approach mentioned in the companion paper [1], in which a researcher assess how student's knowledge compares to **expert ideal knowledge** (correct ideas needed to solve the problem) to assess the quiz physics grade. The grade considers the following subcategories: 1) invoking the principles needed to solve the problem (e.g. in this problem, energy conservation and Newton's second law in the context of circular motion), justifying the choice of principles; and invoking surplus principles for the problem (this subcategory was assessed as negative) 2) applying these principles correctly. We weighed each subcategory as worth 1,2/3,1/2,1/3 or 0 point if +, ++/-, +/-, +/-- or -, respectively, yet the grade also took into account by different weights we assigned when a student did not invoke a principle and thus would necessarily not apply it in order to prevent "double penalizing".

As shown in the companion paper [1] we differentiated the researcher's judgment of the students' self-diagnosis into physics and presentation grade.

### Self-diagnosis - Physics

The physics self-diagnosis grade reflected both the **expert ideal knowledge**, as well as the **novice knowledge per se** (what ideas the student believes are needed to solve the problem are reflected in his/her solution and diagnosis – $2^{nd}$ approach mentioned in the companion paper [1]), by reflecting in detail which mistakes students could diagnose. This approach allowed us to better differentiate between students.

Eventually, we did not score the "justification" part, as we realized that no one had justified his or her choice of principles. Therefore, we assume that the students did not think of justification as part of the solution procedure, and by extension, the self-diagnosis procedure.

Tables 2 and 3 present a comparison of students' performance of self-diagnosis (SD) of the physics in the alternative groups. Note that all students made mistakes, thus all of them are included in the analysis.

**TABLE 2.** Grades for students - SD Physics

| | Group B | Group C | Group D |
|---|---|---|---|
| Mean | 0.73 | 0.57 | 0.24 |
| Std. Err. | 0.049 | 0.051 | 0.055 |

**TABLE 3.** P Values, ANCOVA analysis - SD physics

| | Group B | Group C | Group D |
|---|---|---|---|
| Group B | | 0.02 | <0.0001 |
| Group C | | | <0.0001 |

One can see that the grades were not very high (between 0.24 - 0.73 on a 0-1 scale). The ANCOVA analysis shows that all groups differ from each other (p value < 0.05); B does the best, while D does the worst. The findings make sense as B got the maximum support. TA's outline provided problem solution while still requiring thought to examine details. Rubric provided structure with which to understand what was needed. D received the least support. Task D simulates the most common diagnostic context: students referring to the back of the book answer. The finding might suggest that we can't expect much from students in the most common diagnostic context.

Table 4 presents students' performance of self-diagnosis in invoking the correct principles and applying them (i.e. the percentage of students who diagnosed their mistakes out of those who made mistakes in each sub category). We do not include the justification and the invoking surplus principles as they are rarely evident in students' self diagnosis.

**TABLE 4.** Self-diagnosis – physics subcategories

|  | Group B | | | Group C | | | Group D | | |
|---|---|---|---|---|---|---|---|---|---|
|  | + | +/- | - | + | +/- | - | + | +/- | - |
| invoking | 55% | 40% | 5% | 35% | 53% | 12% | 40% | 45% | 15% |
|  | Total: 35% | | | Total: 55% | | | Total: 24% | | |
| applying | 15% | 80% | 5% | 15% | 58% | 28% | 10% | 62% | 29% |
|  | Total: 66% | | | Total: 53% | | | Total: 62% | | |

(Legend: +: correct diagnosis; +/-: partial diagnosis: -: no diagnosis. Total: percentages of students who had mistakes in their quiz regarding some subcategory)

Looking at "invoking," one can obtain that in group B, 55% of those who did not invoke one of the principles or both of them correctly diagnosed this mistake completely. In group D, only 40% diagnosed their mistakes completely, and in group C only 35% did so. In "applying", the diagnosis of the application was harder than the diagnosis of the invoking, as no more than 15%, 15% and 10% (in groups B, C and D respectively) of those who had application mistakes diagnosed them completely. We conclude that it is much easier for students to identify problems in invoking principles than to identify problems in how these principles are applied. We note that the difference in between the groups regarding sub categories is aligned with the difference in the overall self diagnosis performance and can be explained along the same lines: group B received maximal guidance, while group D received minimal guidance.

## Self-diagnosis - Presentation

The presentation self-diagnosis grade is divided into 3 subcategories: a) description, b) planning, and c) checking. Thus, the presentation self-diagnosis grade reflects whether or not they realized correctly if in their solution they a) fully described the problem (e.g. sketched a free body diagram, represented the knowns correctly, etc.), b) made clear an appropriate solution plan (e.g. defined the target and intermediate variables), and c) checked the solution (e.g. checking units, limiting cases, etc.). Tables 5 and 6 present a comparison of students' performance of self-diagnosis of the presentation in the alternative groups.

**TABLE 5.** Grades - self-diagnosis Presentation

|  | Group B | Group C | Group D |
|---|---|---|---|
| Mean | 0.42 | 0.10 | 0.12 |
| Std. Err. | 0.022 | 0.023 | 0.025 |

**TABLE 6.** P Values for ANCOVA analysis: self-diagnosis/presentation

|  | Group B | Group C | Group D |
|---|---|---|---|
| Group B |  | <0.0001 | <0.0001 |
| Group C |  |  | 0.57 |

The ANCOVA analysis (again, taking the quiz physics grade as covariate) shows that regarding the presentation-self-diagnosis grades groups D and C differ from group B (p value < 0.05).

The findings make sense as C and D received the least support for the presentation part. The rubric that group B received (see Fig. 1B in the companion paper [1]) required explicitly to explain what is missing in the problem description (a sketch, known and unknown quantities, target variable), solution construction (set of sub problems defined by intermediate variables looked for and the physics principles used to find them) and checking of the answer.

The findings may suggest that in order to let the students better diagnose their mistakes, a clear instruction requiring the students to provide specific details is needed.

Table 7 presents a between-group comparison of the percentage of students who diagnosed their mistakes regarding the different sub categories of presentation-self-diagnosis.

**TABLE 7.** Presentation diagnosis

|  | Group B | | | Group C | | | Group D | | |
|---|---|---|---|---|---|---|---|---|---|
|  | + | +/- | - | + | +/- | - | + | +/- | - |
| Description | 20% | 41% | 39% | 7% | 32% | 62% | 0% | 43% | 57% |
|  | Total: 67% | | | Total: 69% | | | Total: 67% | | |
| Plan | 34% | 46% | 20% | 16% | 53% | 31% | 6% | 55% | 39% |
|  | Total: 85% | | | Total: 75% | | | Total: 86% | | |
| Check | 39% | 59% | 3% | 0% | 82% | 18% | 0% | 88% | 12% |
|  | Total: 100% | | | Total: 100% | | | Total: 96% | | |

(Legend: +: correct diagnosis; +/-: partial diagnosis: -: no diagnosis. Total: percentages of students who had mistakes in their quiz regarding some subcategory)

One can obtain that there is a difference in the diagnosis of description mistakes between group B on one hand, and between C and D on the other hand: while among the students who had description mistakes, 20% in group B fully diagnosed these mistakes, 7% in group C and no one in group D did so. The percentage of those who were wrong in the description of the problem, however did not diagnose any mistakes there, is 39% in group B, 62% in group C and 57% in group D.

In the planning section, although for all groups the overall the scores are quite low, there is a noticeable difference in the diagnosis of mistakes in planning the solution between group B on one hand, and between C and D on the other hand. While 34% of those who had problems in planning in group B fully diagnosed these mistakes, 16% in group C and 6% in group D did so. The percentage of those who had mistakes in the planning, however did not diagnose any mistakes there, is 20% in group B, 31% in group C and 39% in group D.

An interesting aspect of planning the solution is realizing what the target variable is. In this problem the students were supposed to find the normal force acting on a girl moving in a circular motion at a certain point. This normal force indicates the weight of the girl at this point. Many of the students failed to recognize this, and a very common mistake was to find a new mass for the girl at this point.

64% of the students, who did not correctly identify the target variable in group B, fully recognized this problem when diagnosing. Only 33% in group C and 11% in group D did so.

One can also obtain that in the checking subcategory there is a noticeable difference in the diagnosis of mistakes in checking the solution between group B on one hand, and between groups C and D on the other hand. 39% of those who had mistakes in checking the solution in group B fully diagnosed these mistakes, while no one in groups C and D did so. The percentage of those who were wrong in the way they checked the solution, or did not check it at all, however did not diagnose any mistakes there, is 3% in group B, 18% in group C and 12% in group D.

To conclude, we note that the difference in between the groups regarding sub categories is aligned with the difference in the overall self-diagnosis performance and can be explained along the same lines: group B received maximal guidance, while groups C and D received minimal guidance in the presentation part.

## Self-diagnosis vs. Prior Knowledge

*The second question we address is: How does students' performance of self-diagnosis depend upon their prior knowledge?*

Table 8 presents correlations (for all groups and between groups) between the physics grades and the physics self-diagnosis grades, and between the presentation grades and the presentation self-diagnosis grades.

**TABLE 8:** Correlations for PHYsics: Quiz vs. SD, and correlations for PREsentation: Quiz VS. SD

|  | PHY: Quiz vs. SD | | PRE: Quiz vs. SD | |
| --- | --- | --- | --- | --- |
|  | Correlation | P value | Correlation | P value |
| All | -0.172 | 0.12 | 0.277 | 0.01 |

Focusing on the PHY columns, one notes a marginal negative correlation between the grade of the physics part of the quiz, and the grade of the diagnosis of this part; namely, students who make more mistakes are better at diagnosing them. A possible explanation to a negative correlation would be that students, who did well on the original quiz, because they had fewer mistakes, had less chance tracing them. However, this means that there is some bar for which the students are not able to diagnose above. As reported in table 4, it seems that if such a bar indeed exists it would involve diagnosing the application of the principles that few students were able to perform. Such results would suggest that students who did poorly on the quiz would reduce the gap with student who did well in the quiz. Such hypothesis still needs to be examined.

Focusing on the PRE columns, one notes a positive correlation between the grade of the presentation part of the quiz, and the grade of the diagnosis of this part; Students whose presentation was deficient were worse also at diagnosing it. This means that the initial gap between students who did well and students who did not remained in the presentation part.

## DISCUSSION

We described students' performance in self-diagnostic tasks that involved minimal modeling and feedback. Students' performance was far from perfect. One-third to one-half of the students could recognize when they did not invoke the required principle, yet only 1/10th could recognize when they misapplied the principles. Yet, differences in the scaffolding (i.e. providing a correct solution and a self-diagnosis rubric) affected the resulting diagnosis. The self-diagnosis of the physics part possibly allowed students who initially did poorly in the quiz to reduce the gap with students who initially did better.

## ACKNOWLEDGMENTS

The research for this study was supported by ISF 1283/05 and NSF DUE-0442087.